\documentclass[12pt]{article}

\usepackage{amsmath}
\usepackage{amssymb}
\usepackage{graphicx}
\usepackage{subfigure}

\begin{document}

\ \vskip 1.0 in

\begin{center}
 { \large {\bf Trace Dynamics and a  Non-commutative Special Relativity }}

\vskip 0.2 in

\smallskip

{\large{\bf Kinjalk Lochan and T. P.  Singh}}

\medskip

{\it Tata Institute of Fundamental Research,}\\
{\it Homi Bhabha Road, Mumbai 400 005, India.}\\
{\tt email: kinjalk@tifr.res.in, tpsingh@tifr.res.in}\\
\medskip

\vskip 0.5cm
\end{center}

\vskip 1.0 in

\begin{abstract}

\noindent Trace Dynamics is a classical dynamical theory of noncommuting matrices in which cyclic permutation inside a trace is used to
define the derivative with respect to an operator. We use the methods of Trace Dynamics to construct a noncommutative special relativity.
We define a line-element using the Trace over spacetime coordinates which are assumed to be operators. The line-element is shown to be invariant
under standard Lorentz transformations, and is used to construct a noncommutative relativistic dynamics. The eventual motivation for
constructing such a noncommutative relativity is to relate the statistical thermodynamics of this classical theory to quantum mechanics.

\end{abstract}

\section{Introduction}

\noindent Trace Dynamics is a classical dynamical theory of noncommuting finite-dimensional matrices, developed by Adler
and collaborators \cite{adler}. While the eventual goal of the proposers of that theory was to relate it to quantum
theory, Trace Dynamics [TD] possesses some remarkable properties of its own, and in the present article we will be
 concerned with further development and application of some of these properties.  We will show how one can construct
a special relativistic dynamics for non-commuting coordinate operators, while preserving Poincar\`{e} invariance. This is the first
part of a program to relate quantum theory to a more fundamental theory which treats matter and space-time degrees of freedom as non-commutative,
and yet preserves relativistic invariance.

Trace dynamics is the classical dynamics of $N\times N$ matrices (equivalently, operators) $q_r$, whose elements
can be either even grade [bosonic sector] or odd grade  [fermionic sector] elements of Grassmann numbers.
 The Lagrangian ${\bf L} [\{q_r\},\{\dot{q_r}\}]$ in this program is defined as the trace of a polynomial function of
 the matrices $\{q_r\}$ and
 their time derivatives $\{ \dot{q_r} \}$. The derivative with respect to an operator ${\cal O}$, of the trace
${\bf P}$ of a polynomial $P$ made out of non-commuting operators is defined as follows. Infinitesimal variation in $TrP$ is written and
arranged as
\begin{equation}
\delta TrP = \delta {\bf P} = Tr \frac {\delta Tr P}{\delta {\cal O}}\delta{\cal O} =
Tr\frac {\delta {\bf P}}{\delta {\cal O}}\delta{\cal O},
\end{equation}
with $  {\delta{\bf P}}/{\delta {\cal O}}$ being designated as the {\it trace derivative}.
By proceeding as in ordinary classical mechanics one constructs an action; Lagrangian dynamics is derived from an action principle,
and conjugate momenta $p_r$, a trace Hamiltonian ${\bf H}$ and Hamilton's equations of motion are constructed.
Apart from the trace
 Hamiltonian there are two other important conserved quantities. One is the `trace fermionic number' $N\equiv \sum_F q_r p_r$
 obtained by summing over fermionic variables. The other is the remarkable traceless and anti-self-adjoint Adler-Millard constant
 \cite{adler-millard}
\begin{equation}
\tilde{C}\equiv \sum_B [q_r,p_r]  - \sum_F \{q_r,p_r\}
\label{adler-millard}
\end{equation}
which is a result of the invariance of the Lagrangian under global unitary transformations of the $q_r$ and $p_r$. The subscript $B/F$
 denotes sum over commutators/anti-commutators of bosonic/fermionic matrices.  It is profound that such a
conserved commutator should appear in a classical theory in which the matrices and their commutators/anti-commutators take arbitrary
 values. It is the presence of this matrix-valued Noether charge which makes TD different from ordinary classical mechanics. However, at
the effective level, when we do statistical mechanics with these d.o.f. the statistical average of this charge is expected to throw
some light on the effective commutation relations.

It should be emphasized that the nature of non-commutativity in Trace Dynamics and in the discussion below is completely different from that in conventional applications of non-commutative geometry to space-time structure. In the latter, it is natural to specify space-time commutation relations, generally on the Planck scale, at the kinematical and the dynamical level. One would then look for consequences on the dynamics of the implementation of such noncommutativity. On the contrary, Trace Dynamics as such is a {\it classical} dynamical theory, so there are no definite non-commutation properties.
Non-commutativity [by which is meant the standard commutation relations of quantum theory] emerges only at the statistical level, after one has constructed a statistical thermodynamics of the underlying classical theory. {\it At no stage is there a space-time non-commutativity of the kind considered in conventional physical applications of non-commutative geometry}. The role of arbitrary non-commutation relations in the present analysis is to allow for the existence of the
Adler-Millard charge, of which there is no analog in non-commutative space-time physics.

One can legitimately ask as to why one should one try to develop a noncommutative spacetime framework different from the standard one. The answer lies in the motivation for Adler’s original theory of Trace Dynamics. In Adler’s work, the motive is to derive quantum theory from first principles, starting from an underlying theoretical framework, without having to impose quantum commutation relations in an ad hoc manner on a classical Newtonian theory of mechanics. This underlying framework is Trace Dynamics, which by itself is a classical theory in the sense that although the fundamental configuration variables and their conjugate momenta all non-commute with each other, these  commutators take arbitrary values, unlike in quantum theory. One then develops a statistical thermodynamics of the classical mechanical theory of Trace Dynamics, and it is shown that as a consequence of certain Ward identities [which are a direct consequence of the equipartition of the Adler-Millard charge] one obtains quantum dynamics and the standard quantum commutation relations in the thermodynamic approximation. Furthermore, by considering Brownian motion corrections around this thermodynamic approximation, it is demonstrated how state vector reduction and the Born probability rule emerge, and how the theory makes contact with phenomenological proposals for stochastic modifications to Schrodinger dynamics. This development is discussed in detail in the book of Adler \cite{adler}.  We emphasize that in this work, an external classical space-time is taken as given, as part of a classical spacetime geometry.

It can be argued that, not only should one derive quantum dynamics from an underlying theory, one should also remove the dependence of quantum theory on an external classical space-time. In other words, there should exist an equivalent reformulation of quantum theory which does not refer to a classical space and more importantly classical time \cite{singh}.  We propose to develop such a programme using the methods of Trace Dynamics. As a first step we will construct a `Trace Dynamics’ analog in which the space-time degrees of freedom all non-commute with each other. This construction, which we call a noncommutative special relativity, is the content of the present paper. The various space-time commutation relations take arbitrary values. We demonstrate the existence of a generalized Adler-Millard charge, and describe the phase space structure of the theory. In further work, we will construct a statistical thermodynamics of this classical theory, thereby obtaining a generalised quantum dynamics with operatorized space-time. Consideration of Brownian motion corrections to this generalised quantum dynamics is expected to yield a quantum theory on a classical space-time background. An overview of this programme is given in \cite{singhfqxi}.

We now describe the construction of a generalised special relativistic space-time in which the space-time coordinates have arbitrary valued commutators.

\section{Noncommutative Space-time in Trace Dynamics Approach}

In TD, a background space-time is a given, and it is made up of the usual space-time coordinates
$({\bf x},t)$ of a Minkowski space-time, and dynamics obeys Poincar\`{e} invariance. Here, we ask if the techniques of TD can be used to
 raise the coordinates $({\bf x}, t)$ to the level of operators [equivalently, noncommuting matrices], and a non-commutative classical
 dynamics constructed, while still preserving the metric under certain space-time diffeomorphisms (Poincar\`{e} like invariance). The physical motivation for wanting to do so will be described at the end of the article.

Consider a set of four non-commuting finite dimensional matrices [equivalently operators] $(\hat{t}, \hat{x}, \hat{y}, \hat{z})$
which generalize space-time coordinates to (not necessarily self-adjoint) operators having arbitrary commutation relations (which are not
fixed and are completely general as of now, we will comment more on it towards the end), and from which a line-element is defined
by taking a trace as follows :
\begin{equation}
ds^{2} = Trd\hat{s}^2\equiv Tr[d\hat{t}^2 - d\hat{x}^2 - d\hat{y}^2 - d\hat{z}^2].
\end{equation}
We will assume $d\hat{s}^2$ to be self-adjoint, so that its trace is real valued.

We ask for the most general linear transformation to another set of operators $(\hat{t}', \hat{x}', \hat{y}', \hat{z}')$ which leaves
the above line-element invariant. That is,
\begin{equation}
ds^{2} = Tr[d\hat{t}^2 - d\hat{x}^2 - d\hat{y}^2 - d\hat{z}^2] = Tr[d\hat{t}'^2 - d\hat{x}'^2 - d\hat{y}'^2 - d\hat{z}'^2]
\label{LE}
\end{equation}
under
\begin{eqnarray}
 \hat{t}' &=& A_1\hat{t}+ B_1\hat{x} + C_1\hat{y} + D_1\hat{z}\nonumber\\
 \hat{x}' &=& A_2\hat{t} + B_2\hat{x} + C_2\hat{y} + D_2\hat{z}\nonumber\\
 \hat{y}' &=& A_3\hat{t} + B_3\hat{x} + C_3\hat{y} + D_3\hat{z} \nonumber\\
 \hat{z}' &=& A_4\hat{t} + B_4\hat{x} + C_4\hat{y} + D_4\hat{z}
\label{transform}
\end{eqnarray}
Here ${A_i,B_i,C_i,D_i}$ belong to a graded vector space
\begin{equation}
x=x_0+\sum_i x_i\theta_i+\sum_{i<j} x_{ij}\theta_i\theta_j + ..\hspace{1in}   ,
\end{equation}
where $\{x_0, x_i, x_{ij}, ...\}\in {\cal C}$.
In the above equation $\theta_i$s are anticommuting Grassmann numbers. The inclusion of Grassmann numbers in the transformation is a
 permissible generalization since the operators are not necessarily self-adjoint and assumed to not commute with each other.

The form of the transformation can be arrived at by taking clue from the standard Lorentz transformation for a boost along the
 $x$-axis, and the symmetry between the $y$ and $z$ axis. Thus, specializing from Eqn. (\ref{transform}) we propose
\begin{eqnarray}
 \hat{t}' &=& A\hat{t} + B\hat{x} + \alpha \tilde{C}\hat{y} + \alpha \tilde{C}\hat{z}\nonumber\\
 \hat{x}' &=& A\hat{x} + B\hat{t} + \alpha \tilde{C}\hat{y} + \alpha \tilde{C}\hat{z}\nonumber\\
 \hat{y}' &=& \hat{y} + \tilde{B_3}\hat{x} + \tilde{C_3}\hat{t} +\tilde{D_3}\hat{z} \nonumber\\
 \hat{z}' &=& \hat{z} + \tilde{B_3}\hat{x} + \tilde{C_3}\hat{t} +\tilde{D_3}\hat{y}
\label{transform1}
\end{eqnarray}
where $\alpha$ is an ordinary real number.
By substituting this transformation in (\ref{LE}) and by demanding the invariance of $ds^2=Trd\hat{s}^2$ under these
transformations, it can be shown that\\
 (i) All $A,B,\tilde{C},\tilde{D_i}$s must belong to Grassmannn even sector.\\
(ii) Demanding that $Trd\hat{s}^2$, $Trd\hat{t}^2$,  $Trd\hat{x}^2$, $Trd\hat{y}^2$, $Trd\hat{z}^2$ are real, $\hat{t}, \hat{x}, \hat{y}, \hat{z}$ are all
forced to be either self-adjoint or anti-self adjoint. Furthermore, for them to remain real in a generic transformation and
for invariance of
adjointness type, it is required
of all of them to be of same adjointness type and $A,B,\tilde{C},\tilde{D_i}$s to be Grassmann real.\\
 (iii) $A$ and $B$ are Grassmann even elements which satisfy $A^2-B^2=1$. {\it This relation implies that generically $A$ and $B$
are non-Grassmann real numbers} (this assertion also requires that the transformations form a group).\\
 (iv) $\tilde{C_3} = -\tilde{B_3}\equiv E$ are those Grassmann even elements which have their squares identically zero.\\
 (v) $\tilde{D_3}$ is another Grassmann even element proportional to $E$, i.e. $\tilde{D_3}=\kappa E$ where $\kappa$ is an ordinary
 real number.\\
 (vi) $E=\alpha (B-A)\tilde{C}$, where $\tilde{C}$ is the Grassmann element, and $\alpha$ is real number.
Thus $\tilde{D_3}=\kappa\alpha(B-A)\tilde{C}\equiv \eta \tilde{C}$.\\

The requirement of existence of a symmetry group of these transformations (product of two transfrmations with the above properties must be a transformation with the same properties) further forces the choice $\alpha=0$.
Therefore, one is very rigidly constrained to restrict the transformation to the standard Lorentz transformation
\begin{eqnarray}
 \hat{t}' &=& \frac{1}{\sqrt{1-\beta^{2}}}[\hat{t} -\beta \hat{x}]  \nonumber\\
 \hat{x}' &=& \frac{1}{\sqrt{1-\beta^{2}}}[\hat{x} -\beta \hat{t}]\nonumber\\
 \hat{y}' &=& \hat{y} \nonumber\\
 \hat{z}' &=& \hat{z} .
\end{eqnarray}
with $\beta=-B/A$, and $A^2-B^2=1$, as in ordinary special relativity. We have set $c=1$; here it is a universal constant without any further physical interpretation.  With these choices of parameters the transformation of symmetry is devoid of any Grassmann variable and with this structure we clearly obtain a symmetry group over field of real numbers
$$\left(
\begin{array}{llll}
 \frac{1}{\sqrt{1-\beta^{2}}} &  \frac{-\beta}{\sqrt{1-\beta^{2}}} & 0 & 0\\
 \frac{-\beta}{\sqrt{1-\beta^{2}}} &  \frac{1}{\sqrt{1-\beta^{2}}} & 0 & 0\\
0 & 0 & 1 & 0\\
0 & 0 & 0 & 1
\end{array}
\right).$$
The group is the Lorentz group. As the above defined transformations form a group, we have
\smallskip
{\bf (a)Closure}
$$\left(
\begin{array}{llll}
\frac{1}{\sqrt{1-\beta_1^{2}}} &  \frac{-\beta_1}{\sqrt{1-\beta_1^{2}}} & 0 & 0\\
 \frac{-\beta_1}{\sqrt{1-\beta_1^{2}}} &  \frac{1}{\sqrt{1-\beta_1^{2}}} & 0 & 0\\
0 & 0 & 1 & 0\\
0 & 0 & 0 & 1
\end{array}
\right)
\left(
\begin{array}{llll}
\frac{1}{\sqrt{1-\beta_2^{2}}} &  \frac{-\beta_2}{\sqrt{1-\beta_2^{2}}} & 0 & 0\\
\frac{-\beta}{\sqrt{1-\beta_2^{2}}} &  \frac{1}{\sqrt{1-\beta_2^{2}}} & 0 & 0\\
0 & 0 & 1 & 0\\
0 & 0 & 0 & 1
\end{array}
\right)
=
\left(
\begin{array}{llll}
\frac{1}{\sqrt{1-\beta_3^{2}}} &  \frac{-\beta_3}{\sqrt{1-\beta_3^{2}}} & 0 & 0\\
\frac{-\beta_3}{\sqrt{1-\beta_3^{2}}} &  \frac{1}{\sqrt{1-\beta_3^{2}}} & 0 & 0\\
0 & 0 & 1 & 0\\
0 & 0 & 0 & 1
\end{array}
\right).
$$
where, $\beta_3=\frac{\beta_1+\beta_2}{1+\beta_1\beta_2}$. If we start with a generic matrix as given in (\ref
{transform1}) satisfying the subsequent requirements, requirement of closure ensures that $\alpha=0$.
\smallskip
\newline
{\bf (b)Associativity} \\
This is evident since matrix multiplication is associative.\\
{\bf (c)Identity}
$\beta=0$ marks the identity.\\
{\bf (d)Inverse}\\

For

$$\left(
\begin{array}{llll}
\frac{1}{\sqrt{1-\beta^{2}}} &  \frac{-\beta}{\sqrt{1-\beta^{2}}} & 0 & 0\\
 \frac{-\beta}{\sqrt{1-\beta^{2}}} &  \frac{1}{\sqrt{1-\beta^{2}}} & 0 & 0\\
0 & 0 & 1 & 0\\
0 & 0 & 0 & 1
\end{array}
\right)
$$

the inverse is

$$\left(
\begin{array}{llll}
\frac{1}{\sqrt{1-\beta^{2}}} &  \frac{\beta}{\sqrt{1-\beta^{2}}} & 0 & 0\\
 \frac{\beta}{\sqrt{1-\beta^{2}}} &  \frac{1}{\sqrt{1-\beta^{2}}} & 0 & 0\\
0 & 0 & 1 & 0\\
0 & 0 & 0 & 1
\end{array}
\right)$$

We will show next that it is possible to construct a Poincar\`{e} invariant dynamics.  Dynamics is introduced by first defining the four-vector $\hat{x}^{\mu}=({\bf \hat{x}},\hat{t})$. It is important to note that the coordinate operators have been assumed in general to be neither bosonic or fermionic. For further discussion they will be split into their respective bosonic and fermionic parts, so that a given $\hat{x}^{\mu}$ is henceforth either bosonic or fermionic. We define a  four velocity $\hat{u}^{\mu}=d\hat{x}^{\mu}/ds$ and for a particle a four-momentum $\hat{p}^{\mu}$ which satisfies $Tr[\hat{p}^{\mu}\hat{p}_{\mu}]=m^2$ (some invariant). An action
operator $\hat{S}$ and the Trace Action are introduced :
\begin{equation}
S=Tr\hat{S}=\int ds Tr \hat{\cal L}(\hat{x},\dot{\hat{x}}),
\end{equation}
where dot represents derivative wih respect to the parameter $s$.
Extremization of the action leads to the Lagrange equations for the Trace Lagrangian $L=Tr\hat{\cal L}$
\begin {equation}
 \frac{\partial L}{\partial \hat{x}}-\frac{d}{ds}\frac{ \partial L}{ {\partial \dot{\hat{x}}}} = 0.
\end {equation}

Hence,
\begin {equation}
 \frac{ \delta L}{ {\delta \dot{\hat{x}}}} = \int ds   \frac{\delta L}{\delta \hat{x}} = \frac{\delta}{\delta \hat{x}}\int ds L.
\label{mom}
\end {equation}
If we define $ {\delta L}/{ {\delta \dot{\hat{x}}}}$ as momentum conjugate to $\hat{x}$, then
\begin{equation}
 \hat{P}_x = \frac{\delta S}{\delta \hat{x}}.
\end{equation}
Derivatives with respect to operators are to be understood as trace derivatives.

In classical mechanics, this momentum is tangent to the trajectory in configuration space. Assuming the same to hold here
$ \hat{P}_{x^{\mu}} $ is tangent to the curve drawn in the configuration space of $ x^{\mu}$ co-ordinates. Hence,
$ \hat{P}_{x^{\mu}} =\hat{p}^{\mu} $ and
$$ \hat{P}_{\mu} =\frac{\delta S}{\delta \hat{x}^{\mu}} $$
Therefore, the Hamilton-Jacobi equation of motion is
\begin{equation}
 Tr\left(\left(\frac{\delta S}{\delta \hat{t}}\right)^2 -\left(\frac{\delta S}{\delta \hat{x}}\right)^2 -
\left(\frac{\delta S}{\delta \hat{y}}\right)^2 - \left(\frac{\delta S}{\delta \hat{z}}\right)^2
 \right) = m^2.
\end{equation}
\subsection{Phase Space Structure}
The Hamiltonian analog of the dynamics is obtained by going to phase space and constructing the Hamiltonian
$$ {\cal{H}}= Tr\left( \sum_r \hat{p}_r\dot{\hat{x}}_r -\hat{L} \right) $$
and the Hamilton equations of motion follow by considering the variation
\begin{eqnarray}
 \delta{\cal{H}} &=& Tr \sum_r((\delta\hat{p}_r)\dot{\hat{x}}_r +  \hat{p}_r\delta\dot{\hat{x}}_r) - Tr \sum_r \left(
\frac{\delta L}{\delta \hat{x}_r}\delta \hat{x}_r+\frac{ \delta L}{ {\delta \dot{\hat{x}}_r}}\delta\dot{\hat{x}}\right) \nonumber \\
&=& Tr \sum_r((\delta\hat{p}_r)\dot{\hat{x}}_r -\dot{\hat{p}}_r\delta\hat{x}_r) \nonumber\\
&=& Tr \sum_r(\epsilon_r\dot{\hat{x}}_r\delta\hat{p}_r -\dot{\hat{p}}_r\delta\hat{x}_r)
\end{eqnarray}
where the Trace Hamiltonian is a trace functional of operators ${\hat{x}_r}$, ${\hat{p}_r}$. Thus
\begin{eqnarray}
  \frac{\delta{\cal{H}}}{\delta \hat{x}_r} = -\dot{\hat{p}}_r  \label{HEM1}\\
  \frac{\delta{\cal{H}}}{\delta \hat{p}_r} =\epsilon_r\dot{\hat{x}}_r .\label{HEM2}
\end{eqnarray}
Now, the derivative can be performed element-wise following Adler's scheme and $\epsilon_r =\pm1$ if the element belongs to bosonic/fermionic
sector.
\\

This phase space can be equipped with a Generalized Poisson Bracket. For traced operators ${\bf A}(\hat{x}_r,\hat{p}_r,\tau)$,
${\bf B}(\hat{x}_r,\hat{p}_r,\tau)$ the Generalized Poisson Bracket is
\begin{equation}
 \{{\bf A},{\bf B}\}_{GPB}=Tr\sum_r\epsilon_r\left(\frac{\delta{\bf A} }{\delta \hat{x_r}}\frac{\delta{\bf B} }{\delta \hat{p_r}}-
\frac{\delta{\bf A} }{\delta \hat{p_r}}\frac{\delta{\bf B} }{\delta \hat{x_r}}\right)
\end{equation}
With the help of Hamiltonian equations of motion (\ref{HEM1}),(\ref{HEM2}) one can show that
\begin{equation}
 \frac{d{\bf A} }{d\tau}=\frac{\partial{\bf A} }{\partial \tau}+\{{\bf A},{\bf H}\}_{GPB}
\end{equation}
The genralized Poisson Bracket satisfies the Jacobi-identity \cite{adler}, \cite{ABW}  and the Lie algebras of symmetries of the theory
can be represented as Lie algebras of trace functionals under Generalised Poisson Bracket operation.

It might appear that this Poisson bracket structure is in contradiction with the arbitrary noncommutativity of configuration and momentum variables, as Poisson bracket among configuration d.o.f will vanish. In particular the  Poisson bracket of space and time coordinates is zero, whereas they have an arbitrary noncommutativity amongst themselves. However, there is no contradiction here, because these Poisson brackets are never promoted to commutators, as is done in standard quantum theory. Our treatment of the Poisson bracket in fact is exactly the same as that of Adler in Trace Dynamics [see e.g. Section 1.4 of his book \cite{adler}]. There too,  the Poisson bracket of two noncommuting configuration variables is zero, whereas they have an arbitrary noncommutativity.

\subsection{Example and Comments}
For an exposition of this approach, let us consider, as an example, the Lagrangian
 $$ \hat{L} =\frac{1}{2}m\dot{\hat{x}}^2,$$
\begin{eqnarray}
  \delta \hat{L}  &=& \frac{1}{2}m\delta\dot{\hat{x}}\dot{\hat{x}} + \frac{1}{2}m\dot{\hat{x}}\delta\dot{\hat{x}} \nonumber\\
                  &=& \frac{m}{2} (\epsilon_{\dot{\hat{x}}}+1)\dot{\hat{x}}\delta\dot{\hat{x}}.
\end{eqnarray}
$$ Tr\delta\hat{L} = \delta Tr \hat{L} = Tr\left[\frac{m}{2} (\epsilon_{\dot{\hat{x}}}+1)\dot{\hat{x}}\right]\delta\dot{\hat{x}}. $$
Taking the trace derivative,
\begin{equation}
 \frac{\delta Tr \hat{L}}{\delta \dot{\hat{x}}} =\frac{m}{2} (\epsilon_{\dot{\hat{x}}}+1)\dot{\hat{x}}.
\end{equation}
Hence from the equation of motion
\begin{equation}
\frac{d}{ds}\left(\frac{m}{2}(\epsilon_{\dot{\hat{x}}}+1)\dot{\hat{x}}\right)=0.
\end{equation}
Therefore,
\begin{equation}
\frac{m}{2}(\epsilon_{\dot{\hat{x}}}+1)\ddot{{\hat{x}}}=0.
\end{equation}
The solution to the above equation is $\hat{x} \sim \hat{C}s$ (for even graded elements), with $\hat{C}$ being some matrix valued constant
of motion. \\

If  $\hat{L}$ and ${\cal{H}}$ are constructed using $\hat{x}_r$ and $\dot{\hat{x}}_r$ (or equivalently  $\hat{x}_r$  and
 ${\hat{p}_r}$) only with
$c$-numbers, there is a global unitary invariance which preserves the adjointness property. Moreover, the line element defined above
has
as its symmetry group the Poincar\'{e} group whose generator commutes with generators of global unitary transofrmations. Thus, Poincar\'{e}
invariant theories will have, in complete analogy with the construction of Adler and Millard, a Noether charge corresponding to
global unitary invariance of ${\cal{H}}$. It is given by
\begin{equation}
 Q = \sum_{r\in B}[\hat{x}_r,\hat{p}_r] -\sum_{r\in F}\{\hat{x}_r,\hat{p}_r\} .
\end{equation}

The degrees of freedom now involve the bosonic and fermionic components of the space-time operators as well.
We have thus demonstrated that it is possible to construct a special relativity for noncommuting coordinate operators,
by defining an infinitesimal distance using the Trace function.

We end by explaining the motivation behind such a construction, whose further development is left for future work.
 In Adler's Trace Dynamics, the non-commuting matrices are the degrees of freedom of a classical dynamics, which possesses
the conserved Adler-Millard charge as a result of global unitary invariance of the Lagrangian of the theory. The motivation
for studying such a dynamics comes from quantum theory. One would like to derive quantum theory without having to arrive at it
by quantizing a classical theory,  and one would like to explain wave-function collapse in a quantum measurement by a dynamical
process. This is achieved in TD by constructing a statistical mechanics of the classical matrix theory. In the thermodynamic
 approximation one recovers the quantum Heisenberg equations of motion, the standard commutation relations of quantum theory, and
a relativistic  quantum field theory. Consideration of Brownian motion fluctuations around the equilibrium results in a stochastic
non-linear modification of quantum theory which can explain wave function collapse without violating causality.

It is however a limitation of the analysis if the matter degrees of freedom are treated as noncommuting matrices, whereas space-time
degrees of freedom
 are treated in the usual way, as points of a manifold \cite{singh}. Possibly, a more complete treatment would treat space and time
also as operators [i. e. matrices],
and then derive the classical space-time, and a quantum theory on the classical space-time background, as thermodynamic approximation
to this generalized Trace Dynamics. The construction of a noncommuative special relativity as described above is a first step in
that direction. As the next step we will construct the statistical thermodynamics of the theory and obtain the effective average quantities like
$\langle Q \rangle $. These averages, if equipartitioned, will help in determining the degree of non-commutativity at the emergent level - this will be the subject of future work.

\bigskip

\noindent {\bf Acknowledgements}: It is a pleasure to thank Angelo Bassi and Satyabrata Sahu for discussions. The authors would like to thank the Dipartimento di Fisica Teorica, University of Trieste, for hospitality. This publication was made possible through the support of a grant from the John Templeton Foundation. The opinions expressed in this publication are those of the authors
and do not necessarily reflect the views of the John Templeton Foundation.

\end{document}